\documentclass[12pt]{iopart}
\usepackage{iopams}
\usepackage{bm}
\usepackage{epsfig}
\usepackage{graphicx}
\usepackage{amsfonts}
\usepackage{amssymb}
\usepackage{mathrsfs}
\usepackage{color}
\usepackage{appendix}

\begin{document}
\title{Cooling a charged mechanical resonator with time-dependent bias gate
voltages}
\author{Jian-Qi Zhang$^1$, Yong Li$^{2,+}$ and Mang Feng$^{1,*}$}
\address{$^1$ State Key Laboratory of Magnetic Resonance and Atomic and Molecular Physics,
Wuhan Institute of Physics and Mathematics, Chinese Academy of Sciences,
Wuhan 430071, China}
\address{$^2$ Beijing Computational Science Research Center, Beijing 100084, China }
\ead{$^+$ liyong@csrc.ac.cn, $^*$ mangfeng@wipm.ac.cn}

\begin{abstract}
We show a purely electronic cooling scheme to cool a charged mechanical resonator (MR) down to
nearly the vibrational ground state by elaborately tuning bias gate voltages
on the electrodes, which couple the MR by Coulomb interaction. The key step
is the modification of time-dependent effective eigen-frequency of the MR
based on the Lewis-Riesenfeld invariant. With respect to a relevant idea
proposed previously [Li \textit{et al}., Phys. Rev. A \textbf{83}, 043803
(2011)], our scheme is simpler, more practical and completely within the
reach of current technology.
\end{abstract}

\date{\today }
\maketitle

Micro- and nano-mechanical resonators (MRs) have attracted much research
interest due to their combination of both classical and quantum properties
\cite{V.B.Braginsky,prl-97-237201} together with broad impact on fundamental
researches and applications \cite{Physics-2-40}, such as ultra-sensitive
measurements approaching quantum limit \cite{Physics-2-40,prl-108-120801},
observation of continuous-variable entanglement with mesoscopic objects~\cite%
{pra-84-024301,prl-101-200503}, quantum information processing~\cite%
{prl-105-220501}, and biological sensing \cite{Nat.nanotech-3-501}.

The prerequisite of the research in these aspects is to cool the MRs down to
their ground states in order to suppress detrimental influence from thermal
fluctuations. Up to now, there have been many proposals for cooling the MRs
\cite{Nature-460-724, Nature-432-200, PRB-78-134301, prl-92-075507,
prl-99-093901, prl-99-093902,prl-103-227203,pra-83-043804,
PRB-84-094502,prb-80-144508,prl-108-120602,
prb-76-205302,Nature-432-1002} in either optomechanical or
electromechanical systems, in which the resolved sideband cooling method
enables the vibrational ground-state cooling of MRs. So far, the
ground-state cooling of MRs has been achieved experimentally \cite%
{Nature-464-697, Nature-478-89, NatPhys-5-485, Nature-463-72}.

Very recently, a different approach to fast ground-state cooling of MRs has
been proposed~\cite{pra-83-043804} with time-dependent optical driving in a
three-mirror cavity optomechanical system, in which the effective frequency
of the MR can be changed like an `optical spring' and the MR can be cooled
down to nearly its ground- state under the control of time-dependent
external optical driving fields. But the scheme seems challenging
experimentally due to the requirement for both adiabatic evolution under
Born-Oppenheimer approximation and very strong optical powers.

\begin{figure}[tbph]
\includegraphics[width=6cm]{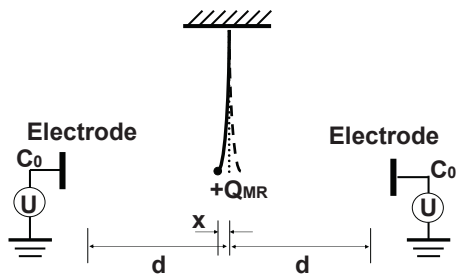}
\caption{Schematic diagram of the system, where a charged MR with (positive)
charge $Q_{MR}$ couples to two identical electrodes via Coulomb interaction.
The two electrodes are distant by $2d$ and $x$ is the deviation of the MR
from the equilibrium position. $C_{0}$ is the capacitance of gate, and $U$
is the tunable bias gate voltage.}
\label{fig B1}
\end{figure}

We present in this work an alternative method using a purely electronic way, which is
simpler but more practical, for cooling a charged MR by two electrodes via
Coulomb interaction. Under the control of bias gate voltages, the MR behaves
as a single-mode harmonic oscillator with its effective frequency tunable
like an `electrical spring'. Our method is somewhat similar to that given in
Ref. \cite{pra-83-043804} but without involvement of Born-Oppenheimer
approximation. Moreover, in contrast to Ref.~\cite{pra-83-043804}, our idea
is more feasible using current experimental techniques due to manipulation
of bias gate voltages. Furthermore, different from the traditional cooling
schemes in electromechanical systems  based on superconductor
qubits \cite{PRB-84-094502,prb-80-144508} or microwave photons \cite%
{PRB-78-134301,prb-76-205302}, our scheme can cool a MR down to its
vibrational ground state by simply tuning the bias gate voltages. To the best of
our knowledge, this is the first cooling scheme with purely electronic way to cool
the MR system without any assistance from auxiliary qubits or additional photons.

As schematically shown in Fig.~\ref{fig B1}, we consider a system where a
charged MR is placed in the middle of two identical electrodes, and couples
electrostatically to these two electrodes. Each electrode takes the charge $%
Q=C_{0}U$ with $C_{0}$ and $U$ being the capacitance and voltage of the bias
gate, respectively. The Coulomb force between the MR and each electrode
varies with the bias gate voltage, and the Coulomb potential is written by
\begin{equation}
V_{c}=kC_{0}UQ_{MR}\left( \frac{1}{d+x}+\frac{1}{d-x}\right) ,  \label{C1-1}
\end{equation}%
where $d$ is the equilibrium separation between the MR center-of-mass
position and the electrodes with $x$ the deviation of the MR from the
equilibrium position, $k$ ($=1/4\pi \varepsilon _{0}$) is the Coulomb
constant in vacuum with $\varepsilon _{0}$ being the vacuum dielectric
constant, and $Q_{MR}$ is the charge of the MR.

After defining $U=U_{0}f(t)$ with the dimensionless time-dependent factor $%
f(t)$ satisfying $|f(t)|<1$, we may reduce Eq.~(\ref{C1-1}), under the
condition of $d\gg x$, to
\begin{equation}
V_{c}\simeq 2kC_{0}U_{0}Q_{MR}f(t)x^{2}/d^{3}.  \label{Vc-approx1}
\end{equation}%
where the term $2kC_{0}U_{0}Q_{MR}f(t)/d$ has been ignored since it commutes
with both the position and the momentum operators of the MR, and makes no
contribution to following discussions.

Supposing the charged MR in the absence of the electrodes takes the
Hamiltonian
\begin{equation}
H_{m}=\frac{1}{2m}p^{2}+\frac{1}{2}m\omega _{m}^{2}x^{2},  \label{C1-21}
\end{equation}%
where $x$ and $p$ are the position and momentum operators of the MR with the
bare eigen-frequency $\omega _{m}$ and the effective mass $m$, we have the
modified Hamiltonian under the control of the bias gate as
\begin{equation}
\begin{array}{lll}
H & = & H_{m}+V_{c} \\
& = & \frac{1}{2m}p^{2}+\frac{1}{2}m\omega _{\mathrm{eff}}^{2}x^{2}%
\end{array}
\label{C1-3}
\end{equation}%
where the effective frequency is $\omega _{\mathrm{eff}}=\omega _{m}\sqrt{%
1+\eta f(t)}$ with $\eta =4k{C_{0}U_{0}Q_{MR}}/(m\omega _{m}^{2}d^{3})$.

Generally speaking, the dynamics of a harmonic oscillator (e.g. MR) governed
by Eq.~(\ref{C1-3}) can be obtained by the method of Lewis-Riesenfeld
invariants \cite{JMP-10-1458,JPB-43-85509, prl-104-063002}. Particularly,
for some special intermediate trajectories, the instantaneous quantum state
of the harmonic oscillator at initial time instant can be the same as that
at the final time instant, although the effective eigen-frequency of the
harmonic oscillator has been changed significantly \cite{JPB-43-85509}. That
is to say, we may have the final mean phonon number of the harmonic
oscillator to be the same as the initial one, but with significant change of
the effective temperature. This idea holds for arbitrary initial state, and
has been used to cool the spatial motion of atoms \cite{prl-104-063002} and
to cool the MR in cavity optomechanical system \cite{pra-83-043804}.

Our present idea is to achieve cooling of the MR by modifying time-dependent
eigen-frequency via tuning the bias gate voltages on external electrodes. We
note that the effective frequency $\omega _{\mathrm{eff}}$ of the MR can be
easily changed by tuning the bias gate voltages $U$. In fact, $\omega _{%
\mathrm{eff}}$ can be much smaller than $\omega_{m}$ under negative $U$.
Moreover, $\omega _{\mathrm{eff}}$ can be even imaginary in order to
accelerate the cooling process~\cite{pra-83-043804,prl-104-063002}.

Specifically, considering some experimentally achievable parameters in
charged MR systems~\cite{Nature-452-72,APS-53-2324, pra-72-041405,
Science-304-74}, such as $k=8.988\times 10^{9}$N$\cdot $m$^{2}/$C$^{2}$, $%
\omega _{m}=2\pi \times 134$ kHz, $m=40$ pg, $d=3.15$ $\mu $m, $U_{0}\equiv
7.00$ V, $Q_{MR}=|e|\sigma _{MR}\times s$, $\sigma _{MR}=1.25\times 10^{13}$%
/cm$^{2}$, $C_{0}=27.5$ nF, and $s=0.08$ $\mu $m$^{2}$, we have $\eta \simeq
1.25\times 10^{7}$, meeting the approximate condition in Eq. (\ref%
{Vc-approx1}).

Similar to Ref.~\cite{pra-83-043804}, our protocol for cooling MR comprises
following two steps. \newline
\newline
\textit{Step I: Decreasing the mean phonon number with the increase of the
MR effective frequency}\newline
\newline
In the absence of bias gate voltages (i.e., $f(t_{o})=0$), we assume the MR
with eigen-frequency $\omega _{\mathrm{eff}}(t_{o})=\omega _{m}$ initially
in a thermal equilibrium state
\begin{equation}
\rho (t_{o})=e^{-\frac{H(t_{o})}{k_{B}T}}/Tr(e^{-\frac{H(t_{o})}{k_{B}T}})
\end{equation}
at temperature $T=20$ mK. So the corresponding mean thermal phonon number is
given by
\begin{equation}
\bar{n}(t_{o})=1/\left\{ \exp [\hbar \omega _{ \mathrm{eff}%
}(t_{o})/k_{B}T]-1\right\}\simeq 3100.
\end{equation}

With an arbitrary trajectory of $\omega _{\mathrm{eff}}(t)$ [or $f(t)$], the
MR is assumed to be in a new thermal state
\begin{equation}
\rho (t_{i})=e^{-H(t_{i})/(k_{B}T)}/Tr \left[ e^{-H(t_{i})/(k_{B}T)}\right],
\end{equation}
at a later time $t_{i}$ with $f(t_{i})=1$, where the effective frequency is $%
\omega_{\mathrm{eff}}(t_{i})\simeq 3500\omega _{m}$ and the mean thermal
phonon number is
\begin{equation}
\bar{n}(t_{i})=\frac{1}{\exp [\hbar\omega_{\mathrm{eff}}(t_{i})/(k_{B}T)]-1}%
\simeq 0.47.
\end{equation}

It is clear that the temperature of the MR at the time $t_{i}$ remains
unchanged compared with the initial one, while the mean thermal phonon
number of the MR is much less than the one at the initial time $t_{o}$ due
to enlargement of the effective frequency $\omega _{\mathrm{eff}}$. The fact
$\bar{n}(t_{i})\simeq 0.47<1$ means that the MR with enlarged effective
frequency has been cooled down to nearly the vibrational ground state.
\newline
\newline
\textit{Step II: Decreasing the MR effective frequency by keeping the low
mean phonon number}\newline
\newline
This step is accomplished via a special trajectory of $f(t)$ evolving to the
final time $t_{f}$ under the control of bias gate voltages satisfying two
conditions: i) The final effective frequency is equal to the bare one (i.e.,
$\omega_{\mathrm{eff}}(t_{f})=\omega _{m}$), implying that the bias gate
voltages are absent again at the final time $t_{f}$ (i.e., $f(t_{f})=0$);
ii) During the cooling process, the mean phonon number remains unchanged
compared with that at time $t_{i}$ (i.e., $\bar{n}(t_{f})=\bar{n}%
(t_{i})\simeq 0.47$). Consequently, we have the MR with the bare frequency
cooled to nearly the vibrational ground state.

The solution to such a special trajectory of $f(t)$ between $t_{i}$ and $%
t_{f}$ governed by Eq. (\ref{C1-3}) follows the Lewis-Riesenfeld invariant
\cite{JMP-10-1458,JPB-43-85509, prl-104-063002}. Using the inverse-invariant
method \cite{JPB-43-85509, prl-80-5469} in Appendix, we obtain the control
function of the trajectory of $f(t)$ by tracing the Lewis-Riesenfeld
invariant,
\begin{equation}
f(t)=\frac{\omega _{\mathrm{eff}}^{2}(t)-\omega _{m}^{2}}{\eta \omega_{m}^{2}%
}=\frac{\omega_{0}^{2}-b(t)^{3}\ddot{b}(t)-\omega _{m}^{2}b(t)^{4}}{\eta
b(t)^{4}\omega_{m}^{2}},
\end{equation}
where the dimensionless function $b(t)$ is defined in Appendix.

\begin{figure*}[tbph]
\includegraphics[width=6cm]{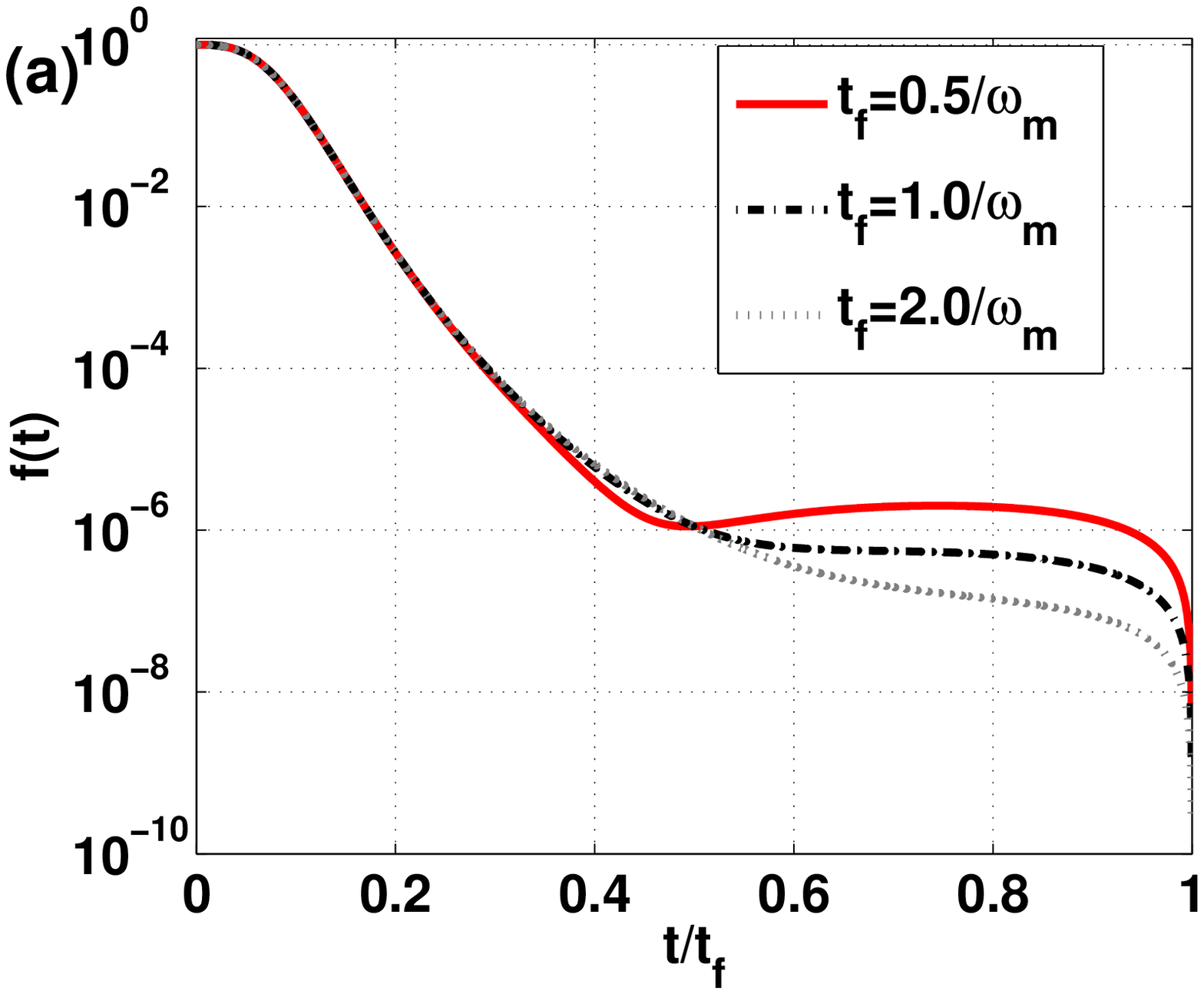} \includegraphics[width=6cm]{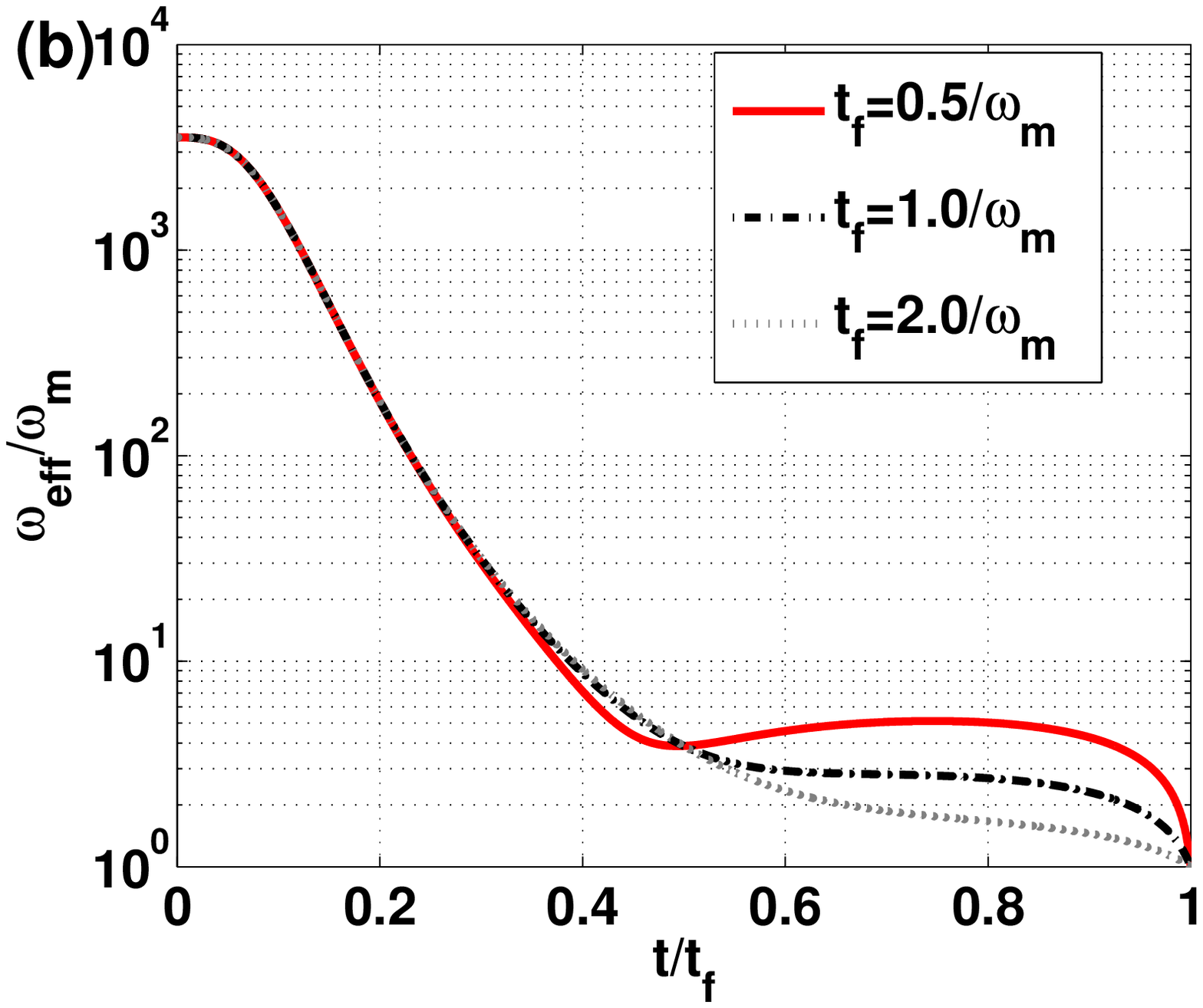} %
\includegraphics[width=6cm]{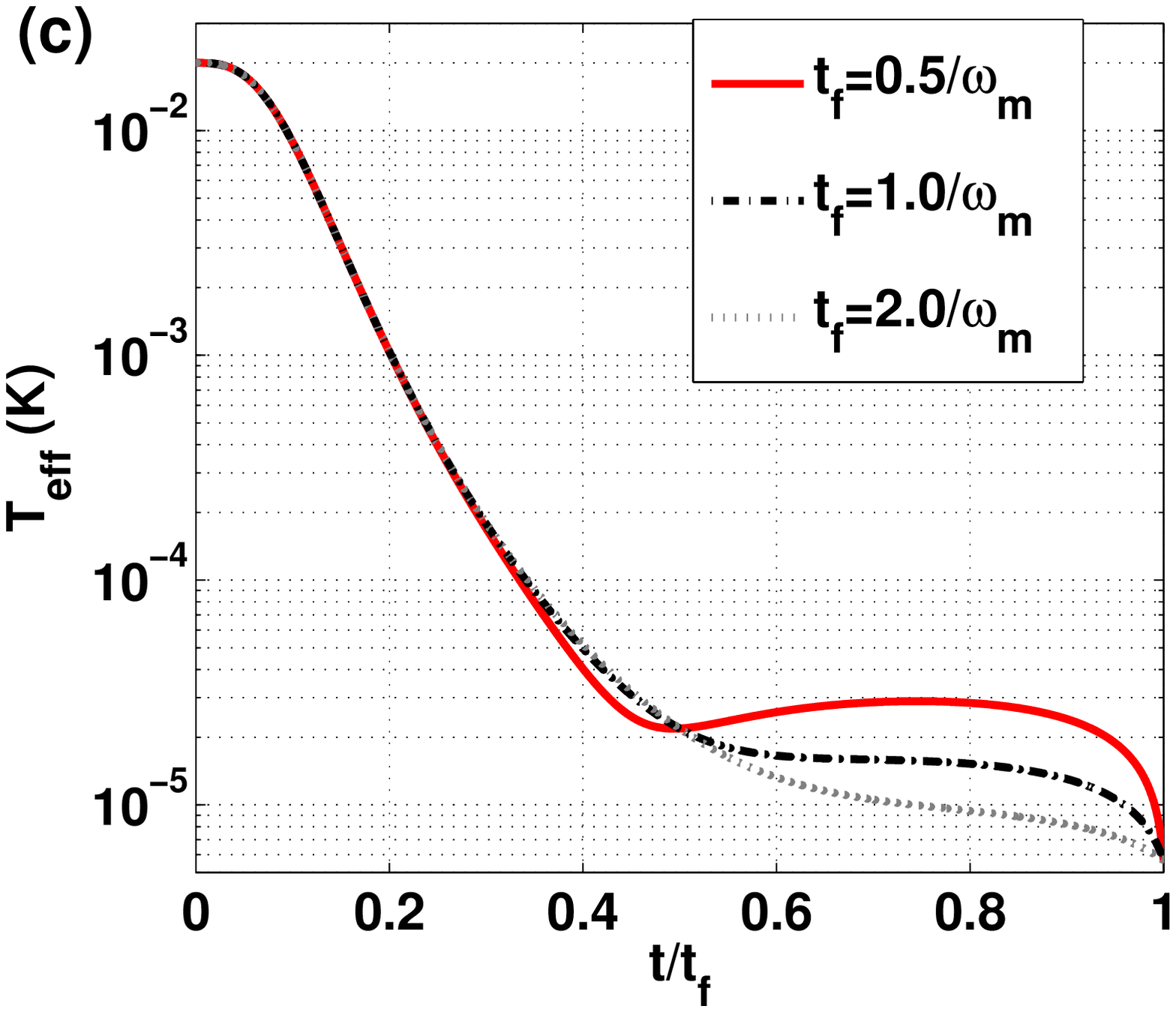}
\caption{(Color online) (a) Time evolution of $f(t)$; (b) Time evolution of
the corresponding effective frequency $\protect\omega _{\mathrm{eff}}$; (c)
Time evolution of the effective temperature of the MR. In each panel, we
consider three cases: $t_{f}=0.5/\protect\omega _{m}$ (red solid line), $%
t_{f}=1.0/\protect\omega _{m}$ (black dotted-dashed line), and $t_{f}=2.0/%
\protect\omega _{m}$ (gray dotted line). The parameter values are taken as $%
\protect\omega _{m}=2\protect\pi \times 134$ kHz; $m=40$ pg; $\protect\sigma %
_{MR}=$ $1.25\times 10^{13}/$cm$^{2}$, $C_{0}=27.52$ nF, $d=3.15$ $\protect%
\mu $m, and $s=0.04\protect\mu $m$^{2}$ from \protect\cite%
{Nature-452-72,APS-53-2324,pra-72-041405,Science-304-74}.}
\label{fig B2}
\end{figure*}

The trajectory of $f(t)$, starting from $f(t_{i}=0)=1$ and ending with $%
f(t_{f})=0$, corresponds to the effective frequency of the MR decreasing
from the end frequency in the first step ($\omega _{\mathrm{eff}%
}(t_{i})\equiv \omega _{0}\simeq 3500\omega_{m}$) to the final frequency $%
\omega _{\mathrm{eff}}(t_{f})=\omega_{m}$. We plot in Fig.~\ref{fig B2} the
control function $f(t)$, the corresponding instantaneous eigen-frequency $%
\omega _{\mathrm{eff}}(t)$ and the effective temperature $T_{eff}$.

\begin{figure*}[tbph]
\includegraphics[width=6cm]{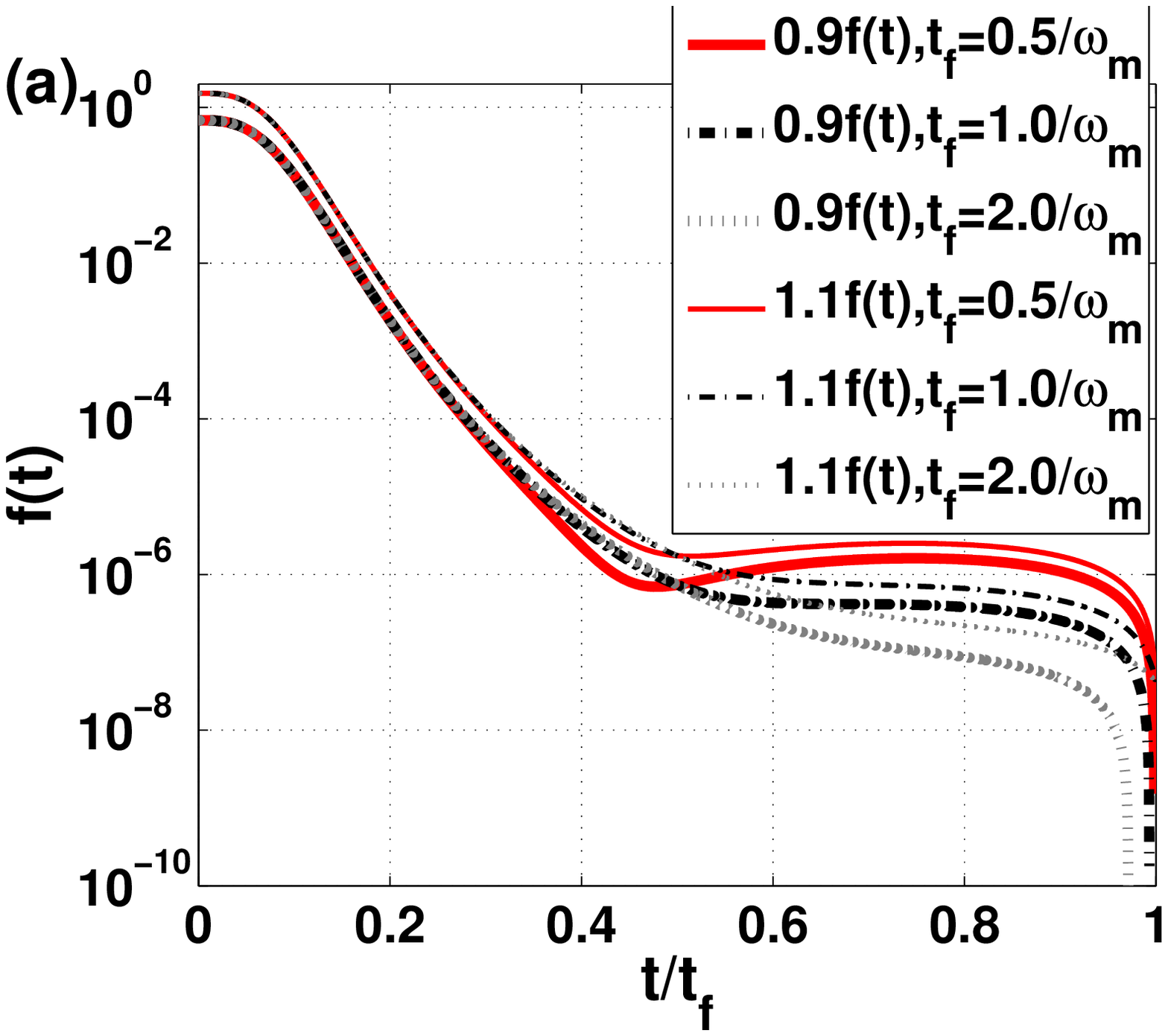} \includegraphics[width=6cm]{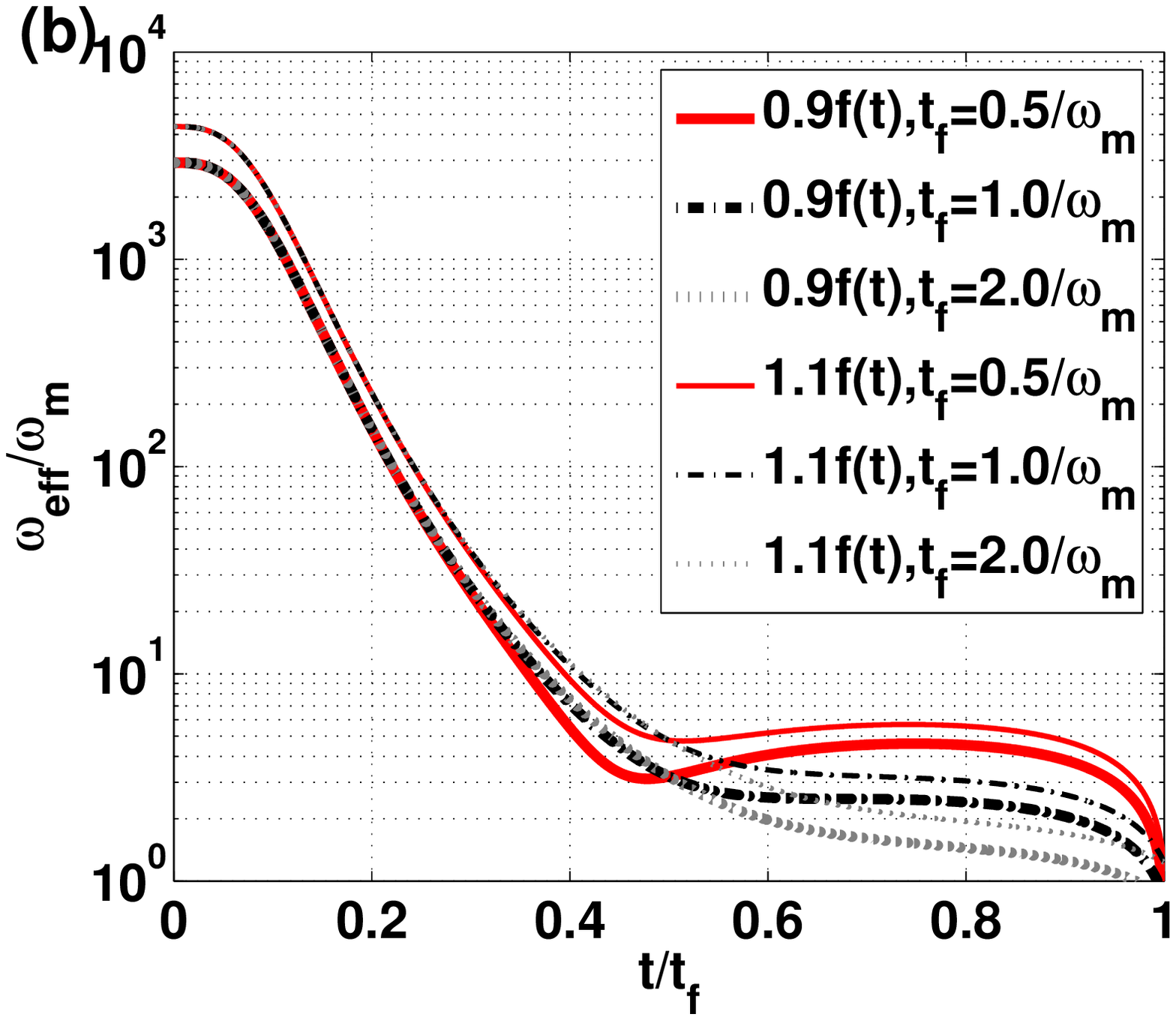} %
\includegraphics[width=6cm]{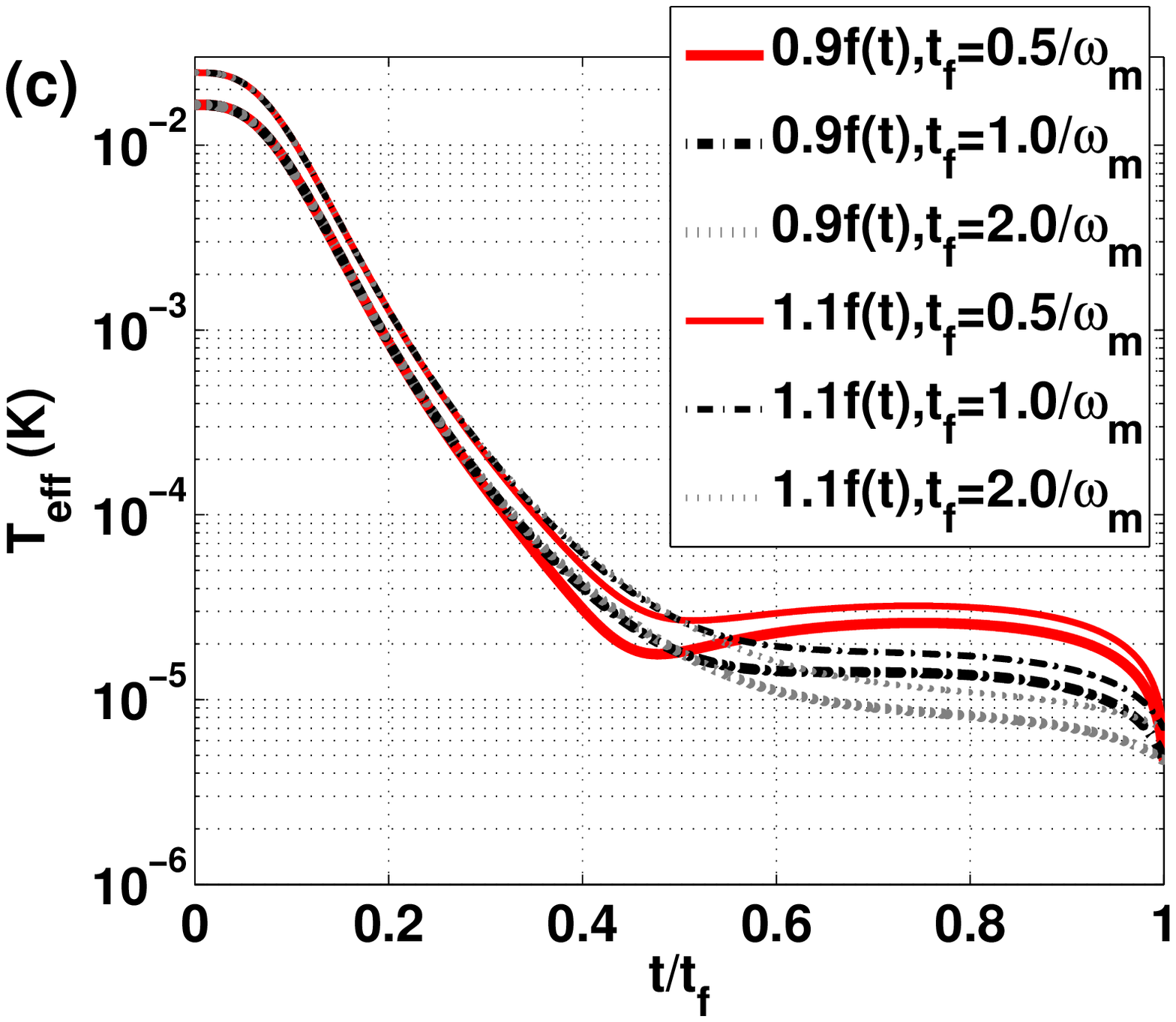}
\caption{(Color online) (a) Time evolution of $f(t)$ with 10$\%$ error (i.e.,
$0.9f(t)=(1-10\%)f(t)$ and $1.1f(t)=(1+10\%)f(t)$); (b) Time evolution of the
effective frequency $\protect\omega_{eff}$ under the different trajectory
of f(t); (c) Time evolution of the effective temperature of the MR under the
different trajectory of f(t). In each plot, we consider two kinds: $%
0.9f(t)=(1-10\%)f(t)$ (thick line), and $0.9f(t)=(1-10\%)f(t)$ (thin line).
Except $f(t)$, other parameters are the same as in Fig.\protect\ref{fig B2}.}
\label{fig B3}
\end{figure*}

Using Eq.~(5), we know that our cooling enables the MR from the initial
temperature $T_{\mathrm{eff}}(t_{i})=T_{\mathrm{eff}}(t_{o})=T=20$ mK down
to the final temperature $T_{\mathrm{eff}}(t_{f})=6$ $\mu $K, as plotted in
Fig.~\ref{fig B2}(c). Here the effective temperature is defined through $%
\bar{n}(t)\equiv 1/\{\exp [\hbar \omega _{eff}(t)/k_{B}T_{\mathrm{eff}%
}(t)]-1\}$.

To estimate the imperfection in the trajectory of $f(t)$, which corresponds
to the experimental error in tunable bias gate voltages on electrodes, we
give the trajectory of $f(t)$ a ten percent fluctuation, i.e., $(1\pm
10\%)f(t)$, as in Fig.~\ref{fig B3}(a), and simulate in Fig.~\ref{fig B3}(b)
and (c) the changes of the instantaneous eigen-frequency $\omega _{\mathrm{%
eff}}(t)$ of the MR and the corresponding effective temperature of the MR.
We find that, although the experimental deviation is as large as $10\%$, the
effective eigen-frequency can still reach $1.23\omega _{m}$ for $+10\%f(t)$ (%
$0.84\omega _{m}$ for $-10\%f(t)$) deviation with the corresponding effective
temperature 7 $\mu $K for $+10\%f(t)$ (5 $\mu $K for $-10\%f(t)$) deviation. In other
words, the MR in such cases can still be cooled down to nearly its
vibrational ground state. Moreover, we note that the final effective temperature 5 $\mu$K
for $-10\%f(t)$ deviation (See Fig.~\ref{fig B3}(c)) seems lower than 6 $\mu$K under the ideal condition.
But this is the case of fluctuation, which is
uncontrollable. In addition, in this deviating case, the effective frequency is $0.83\omega_{m}$, rather than the desired
effective frequency $\omega_{m}$.
Therefore, our choice of trajectory in
Fig.~\ref{fig B3}(b) is definitely optimal.

We would like to point out that our scheme is only for an instantaneous
cooling, which is different from the usual sideband cooling scheme
achievable in steady state. This is because the MR is decoupled from
electrodes at the end of our cooling process. So the MR cannot be kept in
low temperature for very long time but be heated again by environment.

However, after the cooling, the MR would not return to the bath temperature suddenly due to its
high Q factor ($>10^{5}$) \cite {Nature-464-697, Nature-478-89, NatPhys-5-485,
Nature-463-72} which makes sure the small decay of the MR. As a result, there should
be a long enough time to finish the scheduled experimental works, such as quantum
information processing ~\cite {prl-105-220501} before the MR is heated to
the bath temperature. Even if the MR is thermalized to the bath temperature,
we may repeat the same cooling process to put the MR down to the vibrational ground  state again.
Therefore, both the cooling process and the experimental work should be done alternately
in the implementation of the scheduled experimental works.
This is similar to that in ion trap system, wherein one carries out the sideband cooling and
the operations of quantum algorithms alternately  \cite{Roos}.

In summary, we have proposed a practical protocol for cooling a MR near
to its vibrational ground state by controlling bias gate voltages on two
nearby electrodes. Using the achievable experimental parameters and
considering experimental imperfection, we have shown the feasibility of the
proposal by currently available technology. As a result, our work is not only
a practical cooling scheme, but also very promising to verify the
Lewis-Resisenfeld invariant method experimentally.

\appendix*

\section*{Appendix}

We present below how to obtain the special trajectory of $f(t)$ between $%
t=t_{i}$($=0$) and $t=t_{f}$ by means of the Lewis-Riesenfeld invariant.
Based on the inverse-invariant method \cite{JPB-43-85509,prl-80-5469}, Eq.~(%
\ref{C1-3}) satisfies the time-dependent invariant~\cite{JMP-10-1458},
\begin{equation}
I=\frac{1}{2}[\frac{x^{2}}{b(t)^{2}}m\omega _{0}^{2}+\frac{\Lambda ^{2}}{m}],
\label{C1-4}
\end{equation}
where $\omega _{0}=\omega _{\mathrm{eff}}(t_{i})$ for simplicity, $\Lambda
\equiv b(t)p-m\dot{b}(t)x$ takes the role of the momentum conjugate to $%
x/b(t)$, and the dimensionless real function $b(t)$ follows
\begin{equation}
\ddot{b}(t)+\omega _{\mathrm{eff}}^{2}(t)b(t)=\frac{\omega _{0}^{2}}{
b(t)^{3}}.  \label{C1-5}
\end{equation}
According to the boundary conditions at time $t_{i}$ and the finial time $%
t_{f}$, we have
\begin{equation}
\begin{array}{ccc}
b(t_{i})=1, & \dot{b}(t_{i})=0, & \ddot{b}(t_{i})=0, \\
b(t_{f})=\chi , & \dot{b}(t_{f})=0, & \ddot{b}(t_{f})=0,%
\end{array}
\label{C1-6}
\end{equation}
with $\chi =\sqrt{\omega _{0}/\omega _{\mathrm{eff}}(t_{f})}$. So the
simplest polynomial $b(t)$ between $t_{i}$ ($=0$) and $t_{f}$ is given by
\begin{equation}
b(t)=6(\chi -1)s^{5}-15(\chi -1)s^{4}+10(\chi -1)s^{3}+1  \label{b(t)}
\end{equation}
with $s=t/t_{f}$.

\ack
The work is supported by the National Fundamental Research Program of China
(Grant Nos. 2012CB922102, 2012CB922104, 2009CB929604 and 2007CB925204) and
the National Natural Science Foundation of China (Grant Nos. 10974225,
60978009 and 11174027).

\end{document}